# Photonic RF and microwave reconfigurable filters and true time delays based on an integrated optical Kerr frequency comb source

XINGYUAN XU,[1,8] JIAYANG WU,[1,8] THACH G. NGUYEN,[2] MEHRDAD SHOEIBY,[2] SAI T. CHU,[3] BRENT E. LITTLE,[4] ROBERTO MORANDOTTI,[5,6,7] ARNAN MITCHELL,[2] AND DAVID J. MOSS[1*]

[1]*Centre for Micro-Photonics, Swinburne University of Technology, Hawthorn, VIC 3122, Australia*
[2]*ARC Centre of Excellence for Ultrahigh-bandwidth Devices for Optical Systems (CUDOS), RMIT University, Melbourne, VIC 3001, Australia*
[3]*Department of Physics and Material Science, City University of Hong Kong, Tat Chee Avenue, Hong Kong, China.*
[4]*State Key Laboratory of Transient Optics and Photonics, Xi'an Institute of Optics and Precision Mechanics, Chinese Academy of Science, Xi'an, China.*
[5]*INRS-Énergie, Matériaux et Télécommunications, 1650 Boulevard Lionel-Boulet, Varennes, Québec, J3X 1S2, Canada.*
[6]*National Research University of Information Technologies, Mechanics and Optics, St. Petersburg, Russia.*
[7]*Institute of Fundamental and Frontier Sciences, University of Electronic Science and Technology of China, Chengdu 610054, China.*
[8]*These authors contribute equally to this paper.*
*\*dmoss@swin.edu.au*

**Abstract:** We demonstrate advanced transversal radio frequency (RF) and microwave functions based on a Kerr optical comb source generated by an integrated micro-ring resonator. We achieve extremely high performance for an optical true time delay aimed at tunable phased array antenna applications, as well as reconfigurable microwave photonic filters. Our results agree well with theory. We show that our true time delay would yield a phased array antenna with features that include high angular resolution and a wide range of beam steering angles, while the microwave photonic filters feature high Q factors, wideband tunability, and highly reconfigurable filtering shapes. These results show that our approach is a competitive solution to implementing reconfigurable, high performance and potentially low cost RF and microwave signal processing functions for applications including radar and communication systems.

## 1. Introduction

Photonic microwave and radio frequency (RF) signal processing [1–3] has attracted great interest for a wide range of applications such as radar and communications due to its very high performance such as broad bandwidths, very low loss, high versatility and reconfigurability, and strong immunity to electromagnetic interference. Many key functions have been realized, such as those based on RF time delays including phased array antennas (PAAs), microwave photonic filters (MPFs), analog-to-digital or digital-to-analog conversion, and arbitrary waveform generation [4–7], as well as RF spectrometers [8–10], high fidelity microwave tone generation [11] and many others. For microwave and RF time delays a diverse range of photonic approaches has been proposed based on dispersive elements such as single-mode fibre [12], dispersion compensating fibre [13] and fibre Bragg gratings [14–15], slow-light devices based on stimulated Brillouin scattering, integrated resonators [16–18], wavelength conversion coupled with chromatic dispersion [19], and many more [20–21]. Approaches based on switch-controlled dispersive recirculating loops [22], fast sweeping lasers [23], and dispersion-tunable media [24] have also been investigated.

Photonic RF and microwave devices featuring delay-line (i.e., transversal) structures require multiple channel RF time delays. Traditionally, this has been achieved via discrete laser arrays [16, 22–24] or FBG arrays [14–15], which, although offering advantages, have resulted in significantly increased complexity, as well as reduced performance due to a limited number of optical wavelengths and other factors. Alternative approaches, including those based on optical frequency comb (OFC) sources [12], can mitigate this problem, although they too can suffer from drawbacks such as the need for cascaded high frequency electro-optic (EO) [2,25–30] and Fabry-Perot EO [31] modulators that in turn require high-frequency RF sources.

Kerr micro-comb sources [32–39], particularly those based on CMOS-compatible platforms featuring a high nonlinear figure of merit [34–36], offer many advantages over discrete laser sources, such as the potential to provide a much higher number of wavelengths, a greatly reduced footprint and complexity, as well as significantly improved performance. In particular, for RF transversal functions the number of wavelengths dictates the available channel number of RF time delays. For PAAs, the number of radiating elements determines the beam-width, and so improved angular resolution can be achieved by enlarging the channel number. Similarly, for MPFs, extending the number of taps (channels) results in an increased filtering quality factor ($Q_{RF}$) and time-bandwidth product.

In this paper, we propose and demonstrate advanced transversal photonic microwave and RF signal processing functions. We report the first multi-channel RF tunable microwave true time delay lines for PAAs based on an integrated on-chip micro-ring resonator (MRR) optical frequency comb source. By generating a broadband Kerr comb with a large number of comb lines, we significantly improve the performance and reduce the size, potential cost, and complexity of the true time delay device. By programming and shaping the optical comb, we show that this device is capable of achieving record high angular resolution and a wide range of beam steering angles with very little beam "squint" (variation in beam steering angle with RF frequency).

Further, we demonstrate highly reconfigurable microwave filters by achieving a range of new functions including low pass, half-band highpass, half-band lowpass, band-stop, Nyquist, and bandpass microwave photonic filters (MPFs). We achieve wide center frequency tunability for the bandpass filters without the need for hardware tuning devices such as tunable delay lines, i..e., by only adjusting the tap weights. Our experimental results agree with theory, verifying the feasibility of our approach towards the realization of high performance, versatile, microwave and RF signal transversal processing functions with potentially lower cost and footprint than other solutions.

The rest of the paper is organized as follows: Section 2 introduces the RF and microwave true time delays based on Kerr combs and their applications to PAAs and MPFs. The details of the MRR for comb generation are described as well. In Section 3, we introduce the PAA based on the RF true time delays and analyze the enhanced performance brought about by the use of Kerr combs. Section 4 introduces the microwave photonic filters based on RF true time delays, and demonstrates the enhancement in Q factor brought about by the use of Kerr combs as well as the versatile filter functions achieved by means of line-by-line comb shaping. Section 5 concludes the paper.

## 2. RF true time delays based on integrated optical frequency combs

Figure 1 shows a diagram of our RF true time delays based on an integrated optical comb source. First, Kerr combs were generated based on an integrated MRR. When the wavelength of the pump light was tuned into one of the MRR resonances, with the pump power high enough to provide sufficient parametric gain, optical parametric oscillation occurred [32-39], ultimately generating Kerr optical combs with nearly equal line spacing (as shown in Fig. 1(a)). Then the Kerr combs were shaped and directed to a Mach-Zehnder modulator (MZM), where replicas of the input RF signal were generated on each wavelength, leading to multiple RF channels (as shown in Fig. 1(b)). Next, time delays were introduced between the RF channels by a dispersive element (as shown in Fig. 1(c)), producing multi-channel RF true time delays for subsequent applications including phased array antennas and microwave photonic filters.

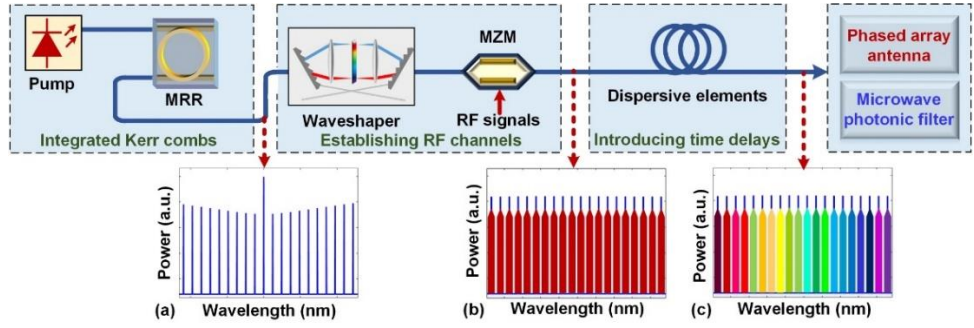

Fig. 1. Schematic diagram of the RF true time delay lines based on an integrated optical comb source. MRR: micro-ring resonator. MZM: Mach-Zehnder modulator.

The MRR used to generate the Kerr optical comb (Fig. 2(a)) was fabricated with high-index doped silica glass using CMOS compatible fabrication processes [34-35]. A scanning electron microscope image of the cross-section of the MRR before depositing the silica upper cladding is shown in Fig. 2(b). The device architecture used a vertical coupling scheme where the gap (approximately 200nm) could be controlled via film growth — a more accurate approach than lithographic techniques [40]. Other advantages of our platform for nonlinear optics include ultra-low linear loss (~0.06 dB·cm$^{-1}$), a sufficient nonlinearity parameter (~233 W$^{-1}$·km$^{-1}$), and most important, negligible nonlinear loss (up to ~25 GW·cm$^{-2}$ [34–35]). After packaging the input and output ports of the device with fibre pigtails, the total insertion loss was ~3.5 dB.

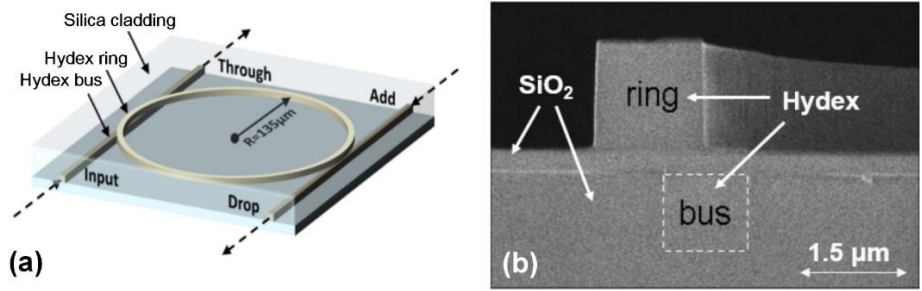

Fig. 2. (a) Schematic illustration of the MRR. (b) SEM image of the cross-section of the MRR before depositing the silica upper cladding.

## 3. High-resolution phased array antenna

Figure 3 shows a diagram of a high-resolution phased array antenna based on our RF true time delays. For the integrated comb source, continuous-wave (CW) light from a tunable laser was amplified by an erbium-doped fibre amplifier (EDFA) to pump the on-chip MRR. Before the MRR, a tunable optical bandpass filter and polarization controller were employed to suppress the amplified spontaneous emission noise and adjust the polarization state, respectively. To avoid resonance drift and maintain the wavelength alignment of the resonances to the pump light, the MRR was mounted on a temperature controller stage.

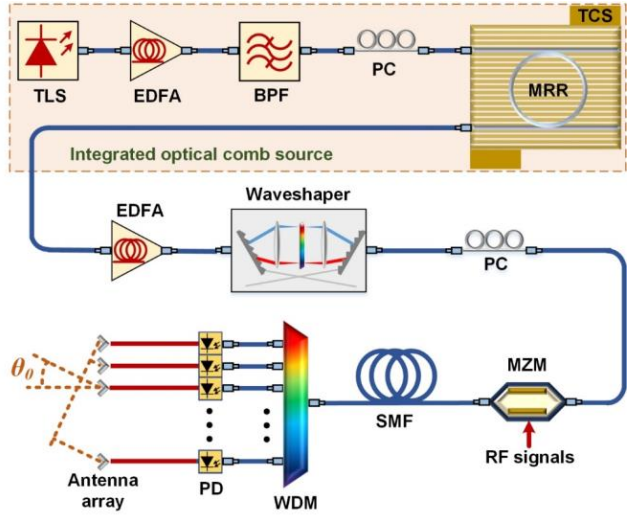

Fig. 3. Schematic diagram of the true time delay device based on an integrated optical comb source. TLS: tunable laser source. EDFA: erbium-doped fibre amplifier. BPF: optical bandpass filter. PC: polarization controller. TCS: temperature controller stage. MZM: Mach-Zehnder modulator. SMF: single mode fibre. WDM: wavelength division multiplexer. PD: photodetector.

Due to the ultra-low loss of our platform, the MRR featured a narrow resonance linewidth (Fig. 4(a)) corresponding to a Q factor of ~1.2 million. The compact integrated MRR had a radius of ~135 μm with a relatively large free spectral range (FSR) of ~1.6 nm (200 GHz, Fig. 4(a)). Such a large FSR enables a potentially large Nyquist zone of ~100 GHz, which is challenging for mode-locked lasers and externally-modulated comb sources [41–42]. By boosting the power of the CW light from the tunable laser source via an EDFA and adjusting the polarization state, multiple FSR mode-spaced combs were first generated, in which the primary spacing was determined by the parametric gain. When the parametric gain lobes became broad enough, secondary comb lines with a spacing equal to the FSR of the MRR were generated via further parametric gain (centred around the different primary comb lines) as well as both degenerate and non-degenerate four wave mixing (FWM). The power threshold for the generation of secondary comb lines was ~500 mW, with our measurements typically performed at roughly 0.8W pump power (in-fibre before coupling on to the chip). Owing to the ultra-high Q factor and small footprint of the MRR, the resulting Type II Kerr optical comb [37, 43] (Fig. 4(b)) was over 200-nm wide, flat over ~32 nm, indicating that in principle, more than 100 channels were available. In practice, however, operation was limited to the C (and potentially L) band due to bandwidth limitations of the waveshaper as well as the EDFA. For these experiments we utilized the C band only, resulting in a maximum of 21 channels. The generated comb was stable in intensity, although it was only partially coherent. Achieving full coherence was not crucial to achieve high performance of our approach. Following generation, the Kerr comb was amplified, shaped by a waveshaper (Finisar 4000s) and directed to a MZM (EOSPACE) in order to establish the RF channels. The output optical signals from the MZM were sequentially delayed by a 2.122-km dispersive SMF, where the dispersion was ~17.4 ps/(nm· km), corresponding to a time delay $T$ of ~59 ps between adjacent RF channels. Finally, individual RF channels were separated by a wavelength division multiplexer (WDM), and then converted back into the RF domain by photodetectors. We note that the pulse shaper and dispersive medium are not limitations to the full integration of the system. InP-based integrated pulse shapers have already been demonstrated [44], and integrated dispersive delays have been achieved in integrated form using various approaches including chirped Bragg gratings [45] and photonic crystals [46]. A promising approach to provide integrated optical time delays is through the use of low loss silicon nitride waveguides [47]. Recently demonstrated integrated beam-formers employed discrete laser arrays [48] or delay line arrays [49] to realize multiple radiating elements. While offering some advantages, these approaches nonetheless suffer from drawbacks including a limited number of channels or radiating elements due to stringent fabrication process requirements. Our approach based on an integrated multi-wavelength source is a highly competitive solution compared with these systems.

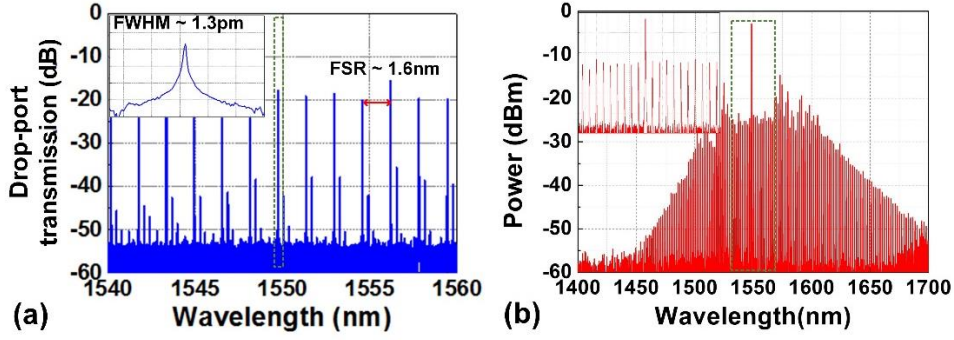

Fig. 4. (a) The drop-port transmission spectrum of the MRR, showing an FSR of ~1.6 nm. Inset shows a resonance at ~1550 nm with full width at half maximum (FWHM) of ~1.2 pm (~150 MHz). (b) Optical spectrum of the generated Kerr comb from 1400 nm to 1700 nm. Inset shows a zoom-in spectrum with a span of ~32 nm.

Antenna arrays consist of uniformly spaced linear radiating elements with a spacing of $d_{PAA}$, where the radiating steering angle $\theta_0$ of the PAA is [50]

$$\theta_0 = \sin^{-1} \frac{c \cdot \tau}{d_{PAA}} \tag{3.1}$$

Here $c$ is the speed of light in vacuum, and $\tau$ is the time delay difference between adjacent radiating elements. From Eq. (1), one can see that the steering angle can be tuned by adjusting $\tau$, which can be accomplished by either employing a dispersion tunable medium [22, 24] or by simply selecting every $m_{th}$ ($m$=1, 2, 3, …) wavelength with the wavelength selective switch (waveshaper). These two approaches for tuning $\tau$ offer either fine tuning steps or a large tuning range, and can be combined together to achieve the best performance for practical applications. In the latter case, $\tau = mT$, where $T$ is the time delay difference between adjacent RF channels and is ~ 59ps. Thus the steering angle is given by

$$\theta_0 = \sin^{-1} \frac{c \cdot mT}{d_{PAA}} \tag{3.2}$$

The corresponding array factor (AF) of the PAA is [50]

$$AF(\theta, \lambda_{RF}) = \frac{\sin^2 \left[ M \pi \left( d_{PAA}/\lambda_{RF} \right) \left( \sin\theta - c \cdot mT/d_{PAA} \right) \right]}{M^2 \sin^2 \left[ \pi \left( d_{PAA}/\lambda_{RF} \right) \left( \sin\theta - c \cdot mT/d_{PAA} \right) \right]} \tag{3.3}$$

where $\theta$ is the radiation angle, $M$ is the number of radiating elements, and $\lambda_{RF}$ is the wavelength of the RF signals. The angular resolution of the PAA is the minimum angular separation at which two equal targets at the same range can be separated, and is determined by the 3-dB beamwidth that can be approximated [51] as $\theta_{3dB} = 102 / M$, which greatly decreases with the number of radiating elements. Our micro-comb source provided up to 21 channels around the pump wavelength over the C-band, thus yielding a high angular resolution for PAA applications.

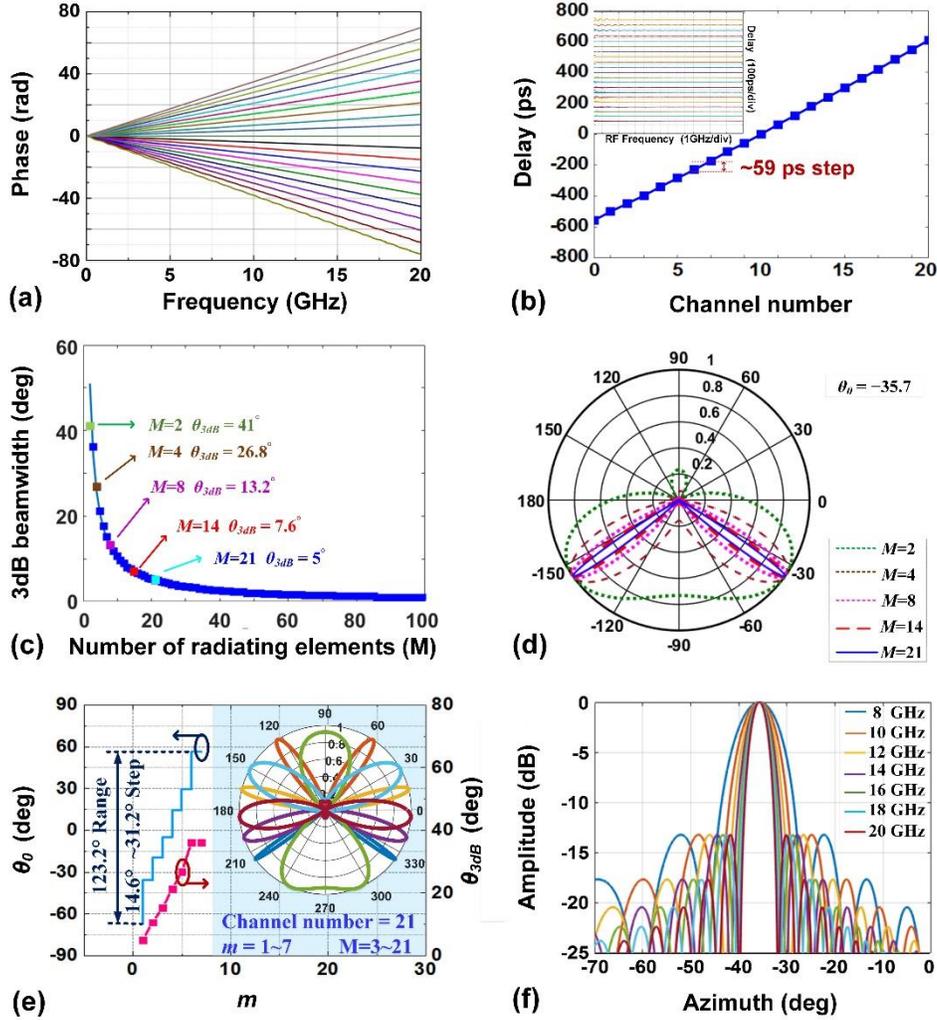

Fig. 5. (a) Measured RF phase responses of the 21-channel true time delay device and (b) corresponding time delays of each channel. Inset in (b) shows the flat delays over a wide RF frequency range. (c) Calculated 3-dB beamwidth of a linear phased array using the true time delay device, as a function of the number of radiating elements (*M*). (d) Calculated array factors (AFs) for various *M* from 2 to 21. (e) Calculated AFs with *m* varying from 1 to 7 and corresponding $\theta_{3dB}$ and $\theta_0$. (f) Calculated AFs with various RF frequencies.

We measured the Kerr comb-based multi-channel RF time delays and conducted a theoretical analysis of the PAA's performance. The RF phase response was measured by a vector network analyser (VNA, Anritsu 37369A, Fig. 5(a)), in which the channel (channel 10) at the central pump wavelength (1548.8nm) was set as the reference. As shown in Fig. 5(b), the phase slopes corresponded to a time delay step of ~59 ps/channel, which matched the calculated time delay difference between adjacent channels (*T*). As seen in Fig. 5(c), increasing *M* would result in a decreased beamwidth and an increased angular resolution for the PAA, which can also be observed (Fig. 5(d)) by plotting the AFs of the PAA as a function of *M*, using Eq. (3).

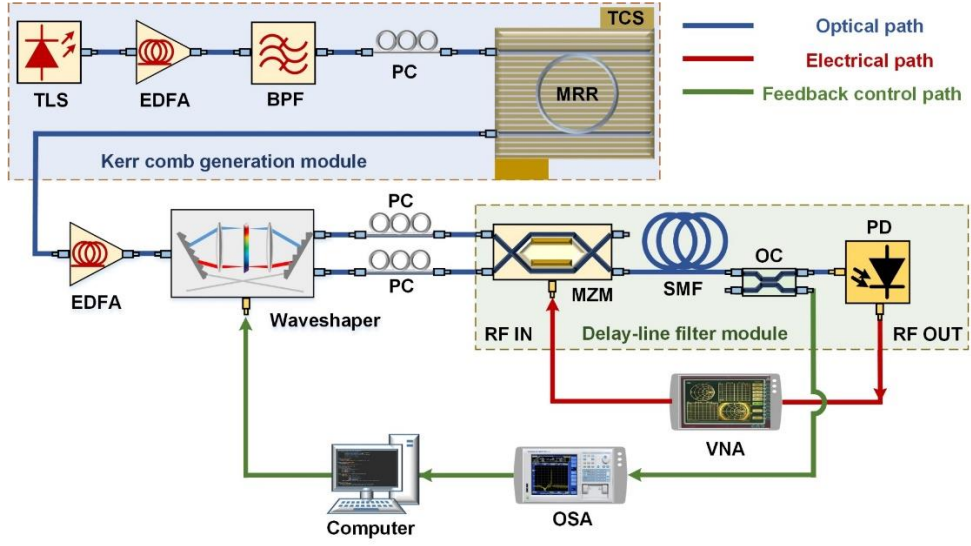

Fig. 6. Schematic diagram of the MPF based on the optical true time delay unit with the integrated optical comb source. TLS: tunable laser source. EDFA: erbium-doped fibre amplifier. PC: polarization controller. BPF: optical bandpass filter. TCS: temperature controller stage. MZM: Mach-Zehnder modulator. SMF: single mode fibre. OC: optical coupler. PD: photodetector. OSA: optical spectrum analyser. VNA: vector network analyser.

In our calculation, $\lambda_{RF}$ is set to the wavelength of a 12 GHz RF signal, $d_{PAA} = \lambda_{RF}/2$, and $\tau = 59$ ps, resulting in a beam steering angle $\theta_0$ of $-35.7°$.

To achieve a tunable beam steering angle for PAAs, we adopted the approach of selecting every $m_{th}$ ($m$=1, 2, 3, …) channel (wavelength) of the comb, and so the time delay between radiating elements could be varied in steps of $T$ (~59 ps in our case). Fig. 5(e) shows calculated AFs as a function of $m$, where $m$ varies from 1 to 7 and $M = \lfloor$Channel Number / $m\rfloor$ (in our case the Channel Number = 21). As m varies, a 123.2° tuning range of the beam steering angle can be achieved with tuning steps from 14.6° to 31.2° (as shown in Fig. 5(e), blue curve). Meanwhile, as $m$ increases, $M$ decreases and leads to a larger 3dB beamwidth $\theta_{3dB}$, thus the angular resolution decreases. Moreover, the (calculated) PAA based on our true time delays could also achieve a wide instantaneous bandwidth without beam squint (the variation of beam steering angle with RF frequency) as seen in Fig. 5(f) where it is clear that the beam steering angle remains the same while the RF frequency varies from 8 to 20 GHz. As a result, the large channel number and high versatility of a PAA based on our device would result in a high angular resolution, a wide tunable range in terms of beam steering angle, and a large instantaneous bandwidth.

## 4. Highly versatile microwave photonic filters

In addition to phased array antennas, our RF time delays are readily applicable to the realization of microwave photonic filters. The transfer function of the MPF can be described as

$$H(\omega) = \sum_{n=0}^{N-1} a_n e^{-j\omega nT} \quad (4.1)$$

where $\omega$ is the angular frequency of the input RF signal, $N$ is the number of taps, $a_n$ is the tap coefficient of the $n_{th}$ tap, and $T$ is the time delay between adjacent channels. The Nyquist frequency of the MPF is given by $f_{Nyquist} = 1/2T$. By imposing calculated tap coefficients $a_n$ on each wavelength channel, reconfigurable MPFs with arbitrary spectral transfer functions could be implemented.

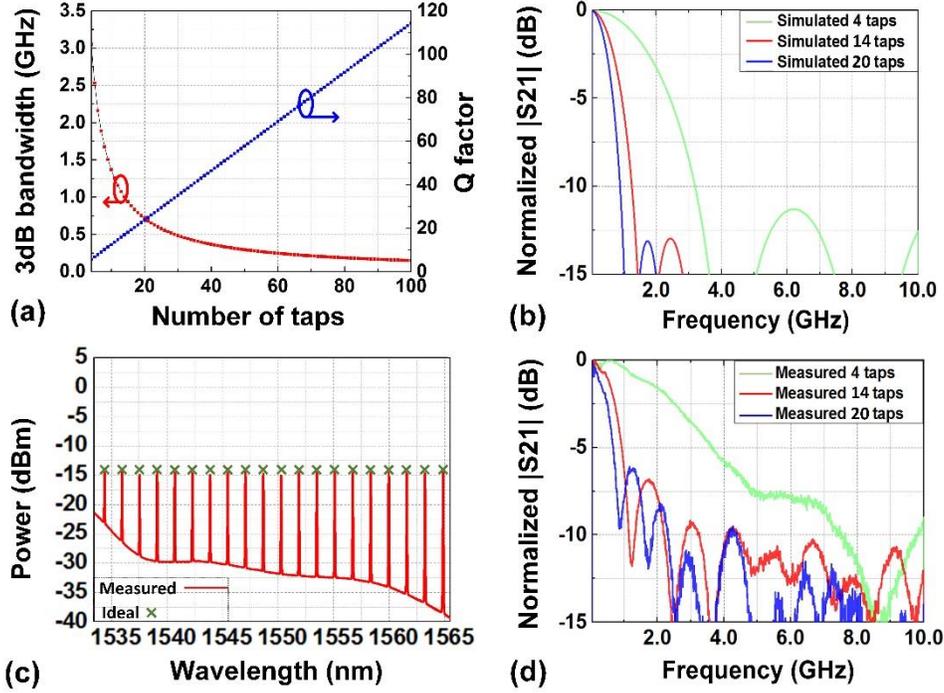

Fig. 7. (a) Correlations between the number of taps and the Q factor, 3dB bandwidth of the all-ones MPF. (b) Simulated RF transmission spectra of an all-ones MPF with different number of taps. (c) Measured optical spectra (red solid) of the shaped optical combs and ideal tap weights (green crossing) for the all-ones MPF. (d) Measured RF transmission spectra of the all-ones MPF with different number of taps.

Figure 6 shows a schematic diagram of the MPFs based on an integrated Kerr optical comb source generated by the on-chip MRR, which was then amplified and fed to the waveshaper. The power of each comb line was manipulated by the waveshaper to achieve appropriately weighted tap coefficients. To increase the accuracy, we adopted a real-time feedback control path to read and shape the power of the comb lines accurately. The comb line powers were first detected by an optical spectrum analyser and then compared with the ideal tap weights. This allowed us to generate an error signal that was fed back into the waveshaper to calibrate the system and achieve accurate comb shaping. The processed comb lines were then divided into two parts according to the algebraic sign of the tap coefficients, and then fed into a 2×2 balanced MZM biased at quadrature. The 2×2 balanced MZM could simultaneously modulate the input RF signal on both positive and negative slopes, thus yielding replicas of the input RF signal with phase and tap coefficients having opposite algebraic signs. The modulated signal produced by the MZM went through ~2.122-km of standard SMF, where the dispersion was ~17.4 ps/(nm·km), corresponding to a minimum time delay $T$ of ~59 ps between adjacent taps (with the channel spacing of the time delay lines set equal to the FSR of the MRR), yielding a Nyquist frequency of ~8.45 GHz for the MPF. We note that the operational bandwidth of the MPF was determined by the Nyquist frequency, which could be easily enlarged by decreasing the time delay and, owing to the large FSR of the compact MRR, could potentially reach over ~ 100

Table 1. Parameters for designing the tunable MPF

| Sampling frequency | 16.9 GHz | Time delay between adjacent taps | 59 fs |
|---|---|---|---|
| Passband width | 0.001 GHz | Nyquist frequency | 8.45 GHz |
| Center frequency ($f_c$) | 2.0005 – 6.4495 GHz | Maximum dynamic range of tap weights | 20 dB |
| Transition bandwidth ($BW_{tr}$) | 0.1 – 2 GHz | Maximum tap number | 20 |

GHz. Finally, the weighted and delayed taps were combined upon detection and converted back into RF signals at the output.

Previous work in micro-comb based MPFs [52, 53] reported bandpass filters where the center frequency was tuned through the use of physical tunable delay lines. In addition to introducing the new filter functions discussed above, we achieve tunability solely by adjusting the tap weights, in turn illustrating the versatility and reach of our approach. Further, we achieve moderately better performance of some parameters such as the quality factor ($Q_{RF}$), of the bandpass filter (the ratio of its 3-dB bandwidth to $T$) as a result of the large number of taps provided by the optical comb source. We note that, to establish a benchmark for the bandpass filter, we used uniform tap coefficients. In Fig. 7(a), we plot the theoretical $Q_{RF}$ factor and 3-dB bandwidth of the "all-ones" MPFs with different tap numbers [6]. Figure 7(b) shows a simulated all-ones MPF with different numbers taps, and as can be seen, the 3-dB bandwidth decreases greatly as the tap number increases. The shaped optical combs (Fig. 7(c)) exhibit a good match between experiment (red solid line) and theory (green crossing), indicating that the comb lines were accurately shaped. The RF response of the all-ones MPF was characterized by a vector network analyser (VNA, Anritsu 37369A) (Fig. 7(d)) confirming the increase in $Q_{RF}$ factor when expanding the tap number from 4 to 20.

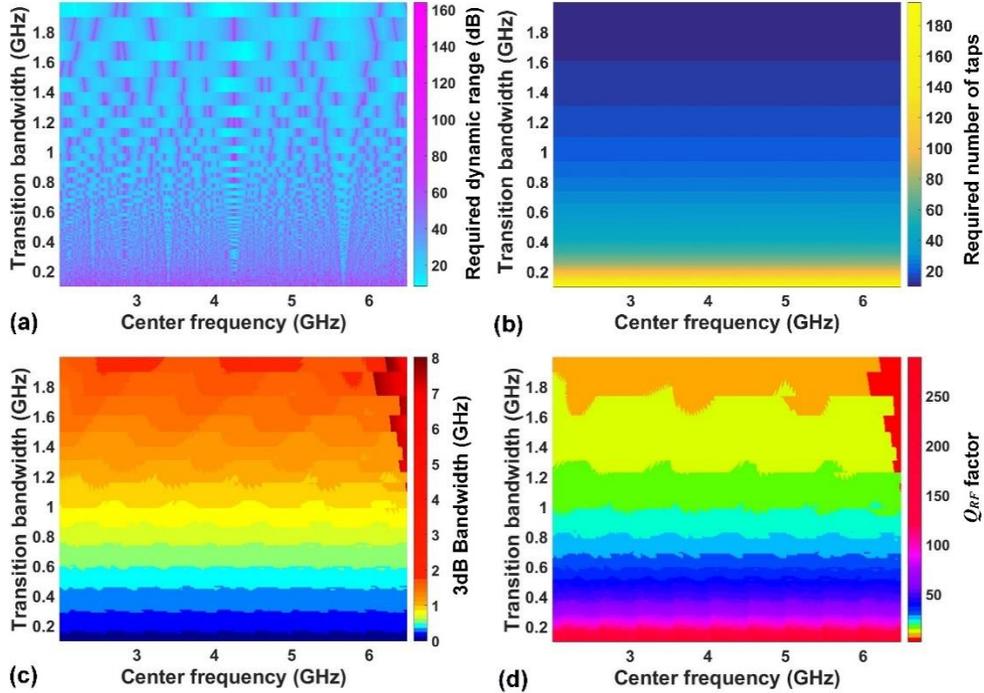

Fig. 8. Calculated (a) required dynamic range of tap weights, (b) required number of taps, (c) 3dB bandwidth and (d) $Q_{RF}$ factor, as functions of the transition bandwidth $BW_{TR}$ and center frequency $f_c$ of the tunable MPF.

**Table 2. Calculated MPF tap coefficients**

| Type | Tap number | Tap coefficients |
|---|---|---|
| Tunable MPF $f_c$=2.135 GHz | 18 | [0.2531, 0.5442, 0.2178, −0.1479, −0.7311, −0.8531, −0.4376, 0.3756, 1, 1, 0.3756, −0.4376, −0.8531, −0.7311, −0.1479, 0.2177, 0.5396, 0.2525] |
| Tunable MPF $f_c$=3.095 GHz | 18 | [−0.3891, −0.3604, 0.3011, 0.6913, 0.3827, −0.6599, −1, −0.1674, 0.9771, 0.9771, −0.1674, −1, −0.6599, 0.3827, 0.6913, 0.3011, −0.3604, −0.3891] |
| Tunable MPF $f_c$=3.956 GHz | 14 | [−0.4733, −0.1911, 0.7966, 0.4053, −0.9784, −0.7763, 1, 1, −0.7763, −0.9784, 0.4053, 0.7966, −0.1911, −0.4733] |
| Tunable MPF $f_c$=4.899 GHz | 19 | [−0.2828, −0.2187, 0.5545, −0.1394, −0.6286, 0.4634, 0.5826, −0.8541, −0.2257, 1, −0.2257, −0.8541, 0.5826, 0.4633, −0.6286, −0.1394, 0.5545, −0.2186, −0.2828] |
| Tunable MPF $f_c$=5.744 GHz | 17 | [−0.2828, −0.2187, 0.5545, −0.1394, −0.6286, 0.4634, 0.5826, −0.8541, −0.2257, 1, −0.2257, −0.8541, 0.5826, 0.4633, −0.6286, −0.1394, 0.5545, −0.2186, −0.2828] |
| Halfband highpass filter | 19 | [−0.0241, 0, 0.0363, 0, −0.0569, 0, 0.1020, 0, −0.3169, 0.5000, −0.3169, 0, 0.1020, 0, −0.0569, 0, 0.0363, 0, −0.0241] |
| Halfband lowpass filter | 15 | [−0.0309, 0, 0.0528, 0, −0.0993, 0, 0.3160, 0.500, 0.3160, 0, −0.0994, 0, 0.0528, 0, −0.0309] |
| Bandstop filter | 21 | [−0.0046, 0.0121, 0, 0.0181, 0.0390, −0.0577, −0.1194, 0.0686, 0.2068, −0.0309, 0.7357, −0.0309, 0.2068, 0.0686, −0.1194, −0.0577, 0.0390, 0.0181, 0, 0.0121, −0.0046] |
| Nyquist filter | 13 | [−0.0361, −0.0347, 0, 0.0685, 0.1530, 0.2228, 0.2500, 0.2228, 0.1530, 0.0685, 0, −0.0347, −0.0361] |

A key attribute of a MPF is its tunability. While in conventional schemes [4, 6] this is generally obtained by physically varying the time delay, here we achieve this by only varying the tap coefficients. This greatly reduces the complexity and instability of the system. We employ the Remez algorithm [54] to determine the tap coefficients for the bandpass MPF with different center frequencies (Table I). Theoretically, the center frequency ($f_c$) and transition bandwidth ($BW_{tr}$) are critical to determining the $Q_{RF}$ factor and the required taps, and so we divided the design of the tunable MPF into two steps. First, the required dynamic range of the generated comb (i. e., the difference between the maximum and minimum power of the comb lines) and the necessary number of taps were calculated as functions of $f_c$ and $BW_{tr}$ (Figs. 8(a–b)). We accounted for practical limitations such as a dynamic range of < 20 dB and a tap number < 20 in order to determine the available sets of $f_c$ and $BW_{tr}$. Secondly, we calculated the bandwidth and $Q_{RF}$ factor (Figs. 8(c–d)) such that the optimized sets of $f_c$ and $BW_{tr}$ were subject to these practical restrictions. As a result, we were able to achieve a tunable MPF with optimized performance by just re-programming the tap weights, without the need for physical tunable delay lines. Thus, the advantages enabled by the use of Kerr combs are clear. In turn, this allowed a large number of wavelengths, providing a large number of taps that resulted in improved $Q_{RF}$ factors of the MPFs.

We selected 5 sets of tap coefficients (Table 2), resulting in 5 different center frequencies for the tunable MPF. Figure 9(a) verifies theoretically that the designed tunable MPFs could achieve acceptable performance. Figure 9(b) shows the quasi-linear correlations between the $Q_{RF}$ factor, reciprocal 3dB bandwidth, and number of taps which is similar to the all-ones MPFs.

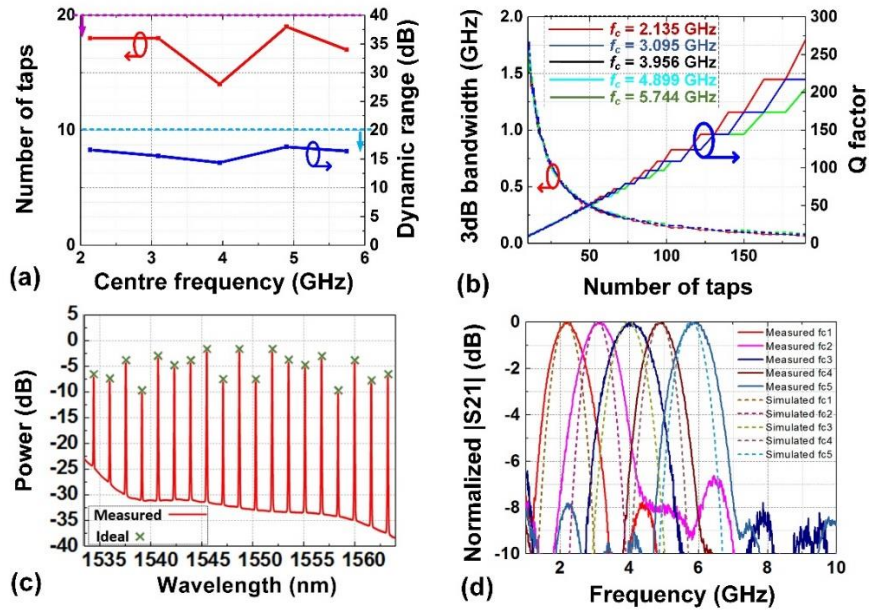

Fig. 9. (a) Parameters of the designed tunable bandpass filter, including number of taps and dynamic range. (b) Correlations between the number of taps and the Q factor, reciprocal 3dB bandwidth of the tunable bandpass filter with different centre frequencies. (c) Measured optical spectra (red solid) of the shaped optical combs and ideal tap weights (green crossing) for the tunable MPF with $f_c$=4.899 GHz. (d) Simulated (dashed) and experimentally measured (solid) RF transmission spectra of a tunable MPF with different center frequencies $f_c$.

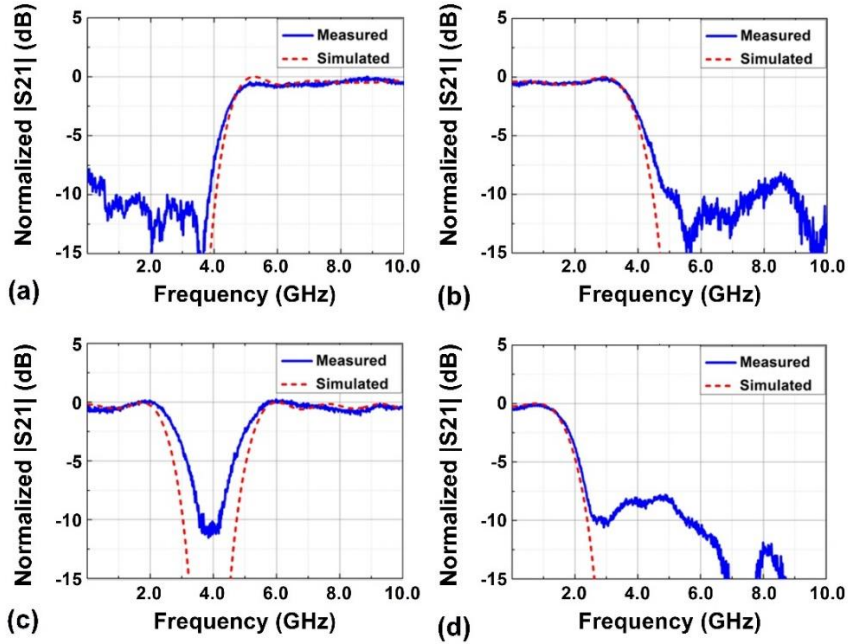

Fig. 10. Measured and simulated RF transmission spectra of (a) halfband highpass filter, (b) halfband lowpass filter, (c) bandstop filter, and (d) Nyquist filter.

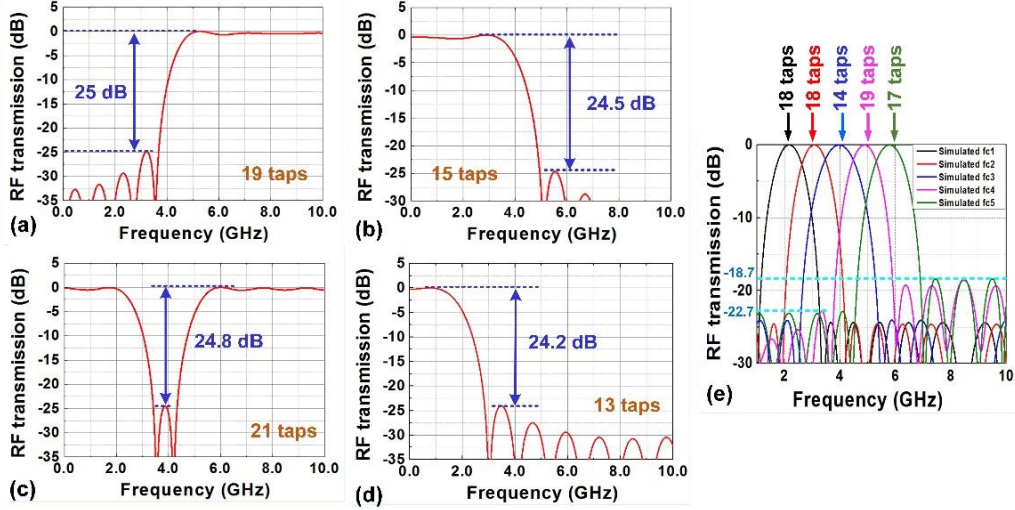

Fig. 11. Simulated RF transmission spectra of (a) halfband highpass filter, (b) halfband lowpass filter, (c) bandstop filter, (d) Nyquist filter, and (e) tunable bandpass filter.

This indicates that increasing the number of taps could significantly improve the $Q_{RF}$ factor or frequency selectivity of the proposed tunable MPF. The shaped optical combs (Fig. 9(c)) show a good match between the measured comb line powers (red solid line) and the calculated ideal tap weights (green crossing), indicating that the comb lines were successfully shaped. The tunability of the MPF (Fig. 9(d)) was demonstrated by adjusting the tap weights according to Table 2, resulting in the center frequencies varying from 25.27% to 57.98% of the Nyquist frequency (8.45 GHz). The Nyquist frequency could also, in principle, be tuned by changing the length of the SMF or the comb spacing to meet diverse requirements in terms of operational bandwidths.

We also realised a range of other transfer functions [55–60] by programming different tap coefficients (Table 2), thus demonstrating a high degree of versatility in terms of center frequency and filter shape. Figure 10 shows four different filter types for microwave signal processing with shapes ranging from a half-band highpass filter, to a half-band lowpass filter, a band-stop filter, and a Nyquist filter. It can be seen that the in-band filtering shape of all four MPFs agrees well with theory, which further confirms the success of the MPF based on our approach.

We note that we did not fully optimize the tap coefficients to yield the highest performing filter shapes, partly because there are many filter parameters that involve trade-offs depending on their priority. The focus of this work was to demonstrate the reach and novelty of our approach, deferring more systematic optimization and design to a future paper. There is a clear roadmap for improvement, particularly for some of the filter parameters such as out of band rejection. Figure 11 shows the theoretical RF transmission spectra of our filters confirming over 20 dB out-of-band rejection with 13 to 21 taps. Further, we have verified that > 50 dB rejection is possible as the number of taps approaches 100 — a feasible target for micro-comb based MPFs. Our device performance was limited by experimental factors such as excess noise from the EDFA, tap weight errors arising during the comb shaping, chirp induced by the MZM, as well as third-order dispersion of the fibre [55]. These can be mitigated by employing low-noise EDFAs, increasing the resolution of the feedback control loop, using a low-chirp MZM, and compensating the third-order dispersion via programming the phase characteristic of the waveshaper.

## 5. Conclusion

We demonstrate high performance, versatile RF and microwave signal processing and filtering functions based on an integrated broadband optical Kerr comb source. The micro-comb source enabled us to obtain significantly enhanced performance in microwave true time delays for phased array antennas, as well as unprecedented versatility for RF filters. We achieve a highly-reconfigurable true RF time delay capable of yielding a high performance phased array antenna with significant angular resolution and very wide range of beam steering angles. We also achieve a wide range of new filtering functions solely by adjusting the channel tap weights (i. e., the power of the Kerr comb lines) through spectral line-by-line shaping. The increased performance, high degree of versatility, and potentially much lower cost of our platform is highly attractive for many microwave and RF applications in signal processing, radar and communications systems.

**Funding**

Australian Research Council Discovery Projects Program (No. DP150104327); Strategic, Discovery and Acceleration Grants Schemes of Natural Sciences and Engineering Research Council of Canada (NSERC); MESI PSR-SIIRI Initiative in Quebec; Canada Research Chair Program; ITMO Fellowship and Professorship Program (grant 074-U 01) of the Government of the Russian Federation; 1000 Talents Sichuan Program in China; Strategic Priority Research Program of the Chinese Academy of Sciences (No. XDB24030000).